\documentclass[aps,prb,manuscript]{revtex4}
\usepackage{graphicx}
\usepackage{epsfig}
\usepackage{color}
\usepackage{bm}

\begin{document}
\draft
\title {Diverse Magnetic Quantization in Bilayer Silicene}

\author{Thi-Nga Do$^{1}$, Po-Hsin Shih$^{2}$, Godfrey Gumbs$^{3}$, Danhong Huang$^{4}$,
Chih-Wei Chiu$^{1}$, and Ming-Fa Lin$^{2}\footnote{Corresponding author: Tel: +886-6-275-7575; Fax: +886-6-74-7995,
{\em E-mail}: mflin@mail.ncku.edu.tw}$}
\affiliation{$^{1}$Department of Physics, National Kaohsiung Normal University, Kaohsiung, Taiwan 82446\\
$^{2}$Department of Physics, National Cheng Kung University, Tainan, Taiwan 701\\
$^{3}$Department of Physics and Astronomy, Hunter College of the City University of New York,
695 Park Avenue, New York, New York 10065, USA\\
$^{4}$US Air Force Research Laboratory, Space Vehicles Directorate, Kirtland Air Force Base,
New Mexico 87117, USA}

\date{\today}

\begin{abstract}
The generalized tight-binding model is developed to investigate the rich and unique electronic
properties of AB-bt (bottom-top)  bilayer silicene under uniform perpendicular electric and magnetic fields.
The first pair of conduction and valence bands, with an observable energy gap, displays unusual
energy dispersions. Each group of conduction/valence Landau levels (LLs) is further classified into
four subgroups, that is, there exist the sublattice- and spin-dominated LL subgroups. The
magnetic-field-dependent LL energy spectra exhibit irregular behavior corresponding to the
critical points of the band structure. Moreover, the electric field can induce many LL anti-crossings.
The main features of the LLs are uncovered with many van Hove singularities in the density-of-states
and non-uniform delta-function-like peaks in the magneto-absorption spectra.
The feature-rich magnetic quantization directly reflects the geometric symmetries,  intra-layer
and inter-layer atomic interactions, spin-orbital couplings, and the field effects. The results of this
work can be applied to novel designs of $Si$-based nano-electronics and nano-devices with enhanced
mobilities.
\end{abstract}
\pacs{PACS:}
\maketitle

\section{Introduction}

Two-dimensional (2D) materials, such as groups IV- and V layered
structures,\,\cite{CHL;NL2012, TND;C2015, PV;PRL2012, TL;NN2015, BM;PE2013, HL;JP2014,JL;RSCA2013,FL;PRL2013,JE;JPCC2015,XW;PCCP2017,HF;APL2014, JYW;RSCA2015, GGG;JMO2016, HD;APL2017, RY;NC2016, GGG;JP2017, PHS;SR2017, Zhu;NM2015, SCC;2017, JYW;PRB2017, GGG;PRB2017} have become main-stream condensed-matter systems since the discovery of graphene in 2004 by mechanical exfoliation. They possess unique geometric properties, nano-scaled thicknesses, a specific lattice
symmetry, a planar/buckled structure, and a stacking configuration. Such systems are verified/predicted
to exhibit diverse physical properties and have many potential device applications. Their Hamiltonians
include complex effects from orbital bonding, spin-orbital coupling (SOC), magnetic fields, electric fields,
and inter-layer atomic interactions. How to solve them becomes one of the basic tasks in solid-state
physics today. This work is mainly focused on  magneto-electronic properties of AB-stacked bilayer silicene.

\medskip
\par

Recently, few-layer silicene with buckled honeycomb lattices have been successfully synthesized
on Ag(111), Ir(111) and ZrBi$_2$ surfaces.\,\cite{PV;PRL2012, TL;NN2015}  According to first-principles calculations,\,\cite{BM;PE2013, HL;JP2014,JL;RSCA2013,FL;PRL2013,JE;JPCC2015,XW;PCCP2017,HF;APL2014}
these  buckled structures  might be meta-stable.   Both AB and AA stackings, being characterized
by the ${(x,y)}$-plane projection, display bottom-top (bt) and bottom-bottom (bb) configurations
on the ${(x,z)}$ plane.\,\cite{JE;JPCC2015} Up to now, the AB-bt and AB-bb configurations have been
confirmed by high-angle annular dark field scanning transmission electron microscopy.\,\cite{RY;NC2016}
The geometric symmetry, the intra-layer and inter-layer atomic interactions, and SOC are expected to
dominate the low-energy physical properties. For example, monolayer silicene presents a slightly displaced
Dirac cone with a narrow direct bandgap (${E_g\sim}10\,$meV) in the presence of SOC.\,\cite{PV;PRL2012}
The low-lying band structures in bilayer silicene become very sensitive to changes in stacking configurations,
such as the stacking-induced indirect gap in AB-bt and semimetal in AA-bb  configurations. The former,
 with the lowest ground state energy, is chosen to be a model system in this paper for studying magnetic
quantization phenomena.

\medskip
\par

The low-energy electronic properties of monolayer silicene are mainly determined by the outer
3$p_z$ orbitals, similar to graphene systems. The perturbation approximation of the ${4\times\,4}$
Hamiltonian could be made around the high-symmetry point (the $\textbf{K}$ point in Fig.\,\ref{Fig1}($a$)),
and then the magnetic quantization follows in a straightforward way. The LL energies are found to
be related to the energy gap and Fermi velocity analytically.\,\cite{CJT;PRB2013} These LLs look
similar to those of monolayer graphene as their magnitudes become much larger than $E_g$. It has been
noticed that the LLs remain doubly degenerate for the spin degree of freedom even with the
SOC,\,\cite{CJT;PRB2013} and the effective-mass approximation becomes too cumbersome for bilayer
silicene with unusual band structures. On the other hand, the generalized tight-binding model has been
developed for solving rigorous Hamiltonians in various condensed-matter systems. It is built on
sub-envelope functions of distinct sublattices (Fig.\,\ref{Fig1}($d$)), in which all the intrinsic
interactions and the external fields can be taken into consideration simultaneously.\,\cite{CYL;PCCP2015}
Specifically, the magnetically quantized energy spectra and wave functions could be evaluated very efficiently
through the exact diagonalization method even for a very large Hamiltonian matrix with complex elements.
This model has been utilized to carry out systematic studies of the magnetic properties of graphene-related
systems.\,\cite{YCH;N2007, YHH;RSCA2015} It has been proven suitable for studying the rich magnetic
quantization phenomena in bilayer silicene with different stacking configurations as well as complicated
interlayer atomic interactions.

\medskip
\par

In this paper, we use the generalized tight-binding model, accompanied with the dynamic Kubo formula and gradient approximation, to investigate the
low-energy electronic and optical properties of AB-bt silicene in the presence of uniform magnetic
($B_z\hat{\mbox{\boldmath{$z$}}}$) and electric ($E_z\hat{\mbox{\boldmath{$z$}}}$) fields. The main
features of the quantized LLs, energy spectra and spatially-oscillating modes are thoroughly
examined, especially for the composite effects arising from intrinsic interactions and external fields.
This work demonstrates that LLs are characterized by the dominating (B$^1$ and B$^2$) sublattices and spin
configurations, leading to four subgroups of conduction/valence LLs.
This LL degeneracy splitting will effectively reduce both the impurity and phonon scatterings and result in enhanced mobilities at
the same time.
The unique LLs are directly reflected in the magneto-optical conductivities with a lot of single, double and twin non-uniform delta-function-like peaks.
The sublattice- and spin-dependent LL energies are confirmed by the calculated DOS, for which the $B_z$-induced energy splitting behaviors could be verified through experiments using scanning tunneling microscopy (STM).
Additionally, the LL energies are found to be tuned easily by an electric field, leading to the crossing and
anti-crossing energy-spectral features which could be selected by the electric field.
Therefore, the use of bilayer silicene, in comparison with monolayer silicene and bilayer graphene, has brought in new opportunity for gate controlled magneto-quantum channel conductance, which is expected to be very useful for novel designs of
$Si$-based nano-electronics and nano-devices.\,\cite{huang1,huang2,huang3}

\section{Method}

\begin{figure}
\centering
{\includegraphics[width=0.8\linewidth]{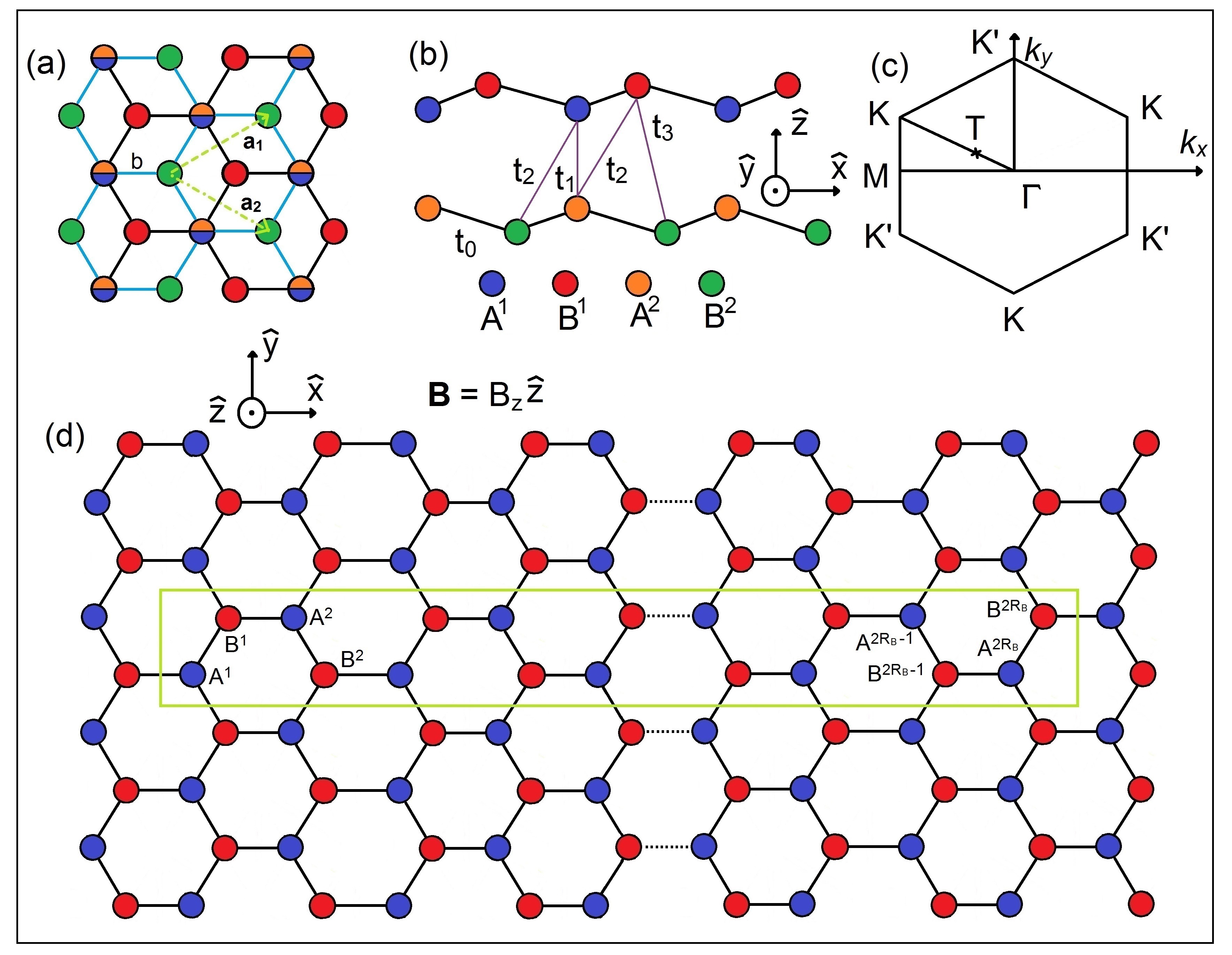}}
\caption{(Color online)  The top view (a) and side view (b) of the atomic structure for bilayer
silicene with intra- and inter-layer atomic interactions. The first Brillouin zone along the high
symmetry points is illustrated in (c), the highly symmetric $\textbf{K}$ ($\bf{K^{\prime}}$) and $\bf{\Gamma}$ points and an extreme one, $\textbf{T}(k_x= 0.1 \times \sqrt{3}, k_y= 0.1)$ (1/$\AA$), are presented.
The enlarged rectangular-shape unit cell under a uniform perpendicular
magnetic field is shown in (d).}
\label{Fig1}
\end{figure}

The generalized tight-binding model has been developed to investigate the feature-rich electronic
properties of AB-bt bilayer silicene arising from intra-layer and inter-layer atomic interactions,
SOC and  the buckled structure. The magnetic and electric fields are simultaneously included in our
calculations.  The AB-bt bilayer silicene, with the low-buckled honeycomb lattice, consists of four
silicon atoms in a unit cell, as depicted in Fig.\,\ref{Fig1}($a$). The two primitive unit vectors,
$\bf{a_1}$ and $\bf{a_2}$, have a lattice constant of $a =3.86\,$\AA.\,\cite{HF;APL2014} Bilayer silicene
acquires four sublattices of ($A^1$, $B^1$) and ($A^2$, $B^2$). For each layer, the two sublattices lie
on two distinct buckling planes with a separation of $l_z =0.46\,$\AA. The $B^1$ and $B^2$ are located
on the higher and lower planes, respectively, for which the inter-layer distance is $2.54\,$\AA. The
buckled angle due to the intra-layer Si-Si bond and the z axis is $\theta =$ 78.3 $^{\circ}$. The
low-energy electronic properties are dominated by the silicon $3p_{z}$ orbitals.
The Hamiltonian built from the tight-binding model includes the intra- and inter-layer atomic interactions, and two kinds of SOCs. The intralayer hopping integral and two kinds of SOCs are similar to monolayer silicene in the previous theoretical prediction.\,\cite{CCL;PRB2011} On the other hand, AB-bt bilayer silicene exhibits the layer-dependent SOCs, a vertical and two non-vertical interlayer atomic interactions, which is absent in monolayer silicene. Such complicated Hamiltonian could be written as

\begin{eqnarray}
\nonumber
H & = &\sum_{m,l}(\epsilon_m^l+U_m^l)c_{m \alpha}^{\dagger l}c_{m \alpha}^{l}
+\sum_{ m,j  , \alpha, l, l^{\prime}} t_{mj}^{ll^{\prime}} c_{m \alpha}^{\dagger l}
c_{j \alpha}^{l^{\prime}}\\
\nonumber
&+&  \frac {i} {3\sqrt{3}} \sum_{ \langle \langle m,j \rangle \rangle, \alpha, \beta, l}
\lambda^{SOC}_{l} \gamma_l v_{mj} c_{m\alpha}^{\dagger l} \sigma_{\alpha\beta}^{z} c_{j\beta}^{l}\\
&-& \frac{2i}{3} \sum_{\langle\langle m,j \rangle \rangle, \alpha, \beta, l} \lambda^{R}_{l} \gamma_l u_{mj}
c_{m\alpha}^{\dagger l} (\vec{\sigma} \times \hat{d}_{mj})_{\alpha\beta}^{z} c_{j\beta}^{l}\ .
\end{eqnarray}
Here, $\epsilon_m^l (A^l,B^l)$ is the sublattice-dependent site energy related to the chemical
environment difference ($\epsilon_m^l (A^l)$ = 0; $\epsilon_m^l (B^l) = -0.12\,$eV).
$U_m^{l} (A^l,B^l)$ is the height-induced Coulomb potential energy arising from a uniform
perpendicular electric field. The $c_{m\alpha}^{l}$/$c_{m\alpha}^{\dagger l}$ operator represents
the annilation/creation of an electronic state with spin polarization $\alpha$ at the $m$-th site of
the $l$-th layer. The atomic interactions in the second term cover the nearest-neighbor intra-layer
hopping integral (${t_0=1.13}\,$eV and three inter-layer hopping integrals due to
($A^1$, A$^2$), ($B^1$, A$^2$) or ($A^1$, B$^2$) and ($B^1$, B$^2$) (${t_1=-2.2}\,$eV, ${t_2=0.1}\,$eV,
${t_3=0.54}\,$eV in Fig.\,\ref{Fig1}($b$)). Specifically, the large inter-layer vertical hopping integral
of $t_1$ induces very strong orbital hybridizations in bilayer silicene. The traditional SOC
(the third term) and the Bychkov-Rashba SOC (the fourth term) take into account the next-nearest-neighbor
pairs $\langle \langle m,j \rangle \rangle$. $\vec{\sigma}$ is the Pauli spin matrix and
$\hat{d}_{mj} = \vec {d}_{mj} / |d_{mj}|$ denotes the unit vector connecting the $m$- and $j$-th lattice sites.
$v_{mj} = \pm 1$ when the next-nearest-neighbor hopping is anticlockwise/clockwise with respect to the
positive z axis. $u_{mj} = \pm 1$ corresponds to the A and B sites, respectively. $\gamma_{l} = \pm 1$ presents the layer-dependent SOCs due to the opposite buckled ordering of AB-bt bialyer silicene. Two kinds of SOCs
appear in the diagonal elements of the Hamiltonian matrix. They are chosen as
${\lambda_1^{SOC}=0.06}\,$eV,  ${\lambda_2^{SOC}=0.046}\,$eV, ${\lambda_1^{R}=-0.054}\,$eV, ${\lambda_2^{R}=-0.043}\,$eV
so that the calculated band structure approaches that from the first-principles
method.\,\cite{XW;PCCP2017}

\medskip
\par

A uniform perpendicular magnetic field can produce an extra Peierls phase in the tight-binding function
through the vector potential $\vec{A}$, leading to an enlarged rectangular-shape unit cell, as illustrated
in Fig.\,\ref{Fig1}($d$). The Peierls phase is characterized by
$G_R =\frac{2\pi}{\phi_0}\int_{R}^{r} \vec{A}.d\vec{l}$, in which $\phi_0 =hc/e$ is the magnetic flux
quantum and $\phi = B_z \sqrt{3}a^2/2$ is the magnetic flux through a hexagon. There are totally
16$R_B$ ($R_B = \phi_0 /\phi$) Si atoms in the enlarged unit cell. The resulting magnetic Hamiltonian,
based on the tight-binding atomic base functions in distinct sublattices, is a 16$R_B$ $\times$ 16$R_B$
Hermitian matrix. The eigenvalues and eigenfunctions of the magnetic Hamiltonian are efficiently numerically solved using the band-like method and the spatial localizations of the magnetic wavefunctions. After the diagonalization of bilayer magnetic Hamiltonian, the Landau level wavefunction, with quantum number n, could be expressed as

\begin{eqnarray}
\Psi (n,\mathbf{k})=\sum_{l=1,2}\sum_{m=1}^{R_B}\sum_{\alpha, \beta} [ A_{\alpha, \beta}^{l,m} (n,\mathbf{k}) |\psi_{\alpha, \beta}^{l,m} (A) \rangle + B_{\alpha, \beta}^{l,m} (n,\mathbf{k}) |\psi_{\alpha, \beta}^{l,m} (B) \rangle
  ].
\end{eqnarray}

Here, $\psi_{\alpha, \beta}^{l,m}$ is the tight-binding function localized at the sublattice-dependent lattice sites, $A_{\alpha, \beta}^{l,m} (n,\mathbf{k})$ ($B_{\alpha, \beta}^{l,m} (n,\mathbf{k})$) is the amplitude on the sublattice-dependent lattice site. Specifically, all the amplitudes in an enlarged unit cell could be regarded as the spatial distributions of the sub-envelope functions on the distinct sublattices. This is because the magnetic distribution width is much larger than lattice constant. Such subenvelope functions provide much information for explaining the interesting LL behaviors, such as the LL quantum number, the localization centers, and the crossing/anticrossing pheonomena. For bilayer silicene, the buckled honeycomb structure, the complex intra- and inter-layer atomic interactions and the significant SOCs need to be fully taken into account in the theoretical model calculations. Such a system is expected to show diverse physical properties under various external fields.\\

The zero-field and magneto optical conductivities are taken into consideration. The optical conductivity can be expressed as $\sigma(\omega) \propto A(\omega)/\omega$, in which, the optical absorption function, A($\omega$), is calculated according to the Fermi golden rule

\begin{eqnarray}
A(\omega) \propto
\sum_{c,v,m,m'} \int_{1stBZ} \frac {d\mathbf{k}}{(2\pi)^2}
 \Big| \Big\langle \Psi^{c} (\mathbf{k},m')
 \Big| \frac{   \hat{\mathbf{E}}\cdot \mathbf{P}   } {m_e}
 \Big| \Psi^{v}(\mathbf{k},m)    \Big\rangle \Big|^2 \nonumber
\end{eqnarray}
\begin{eqnarray}
 \times
Im \Big[      \frac{f(E^c (\mathbf{k},m')) - f(E^v (\mathbf{k},m))}
{E^c (\mathbf{k},m')-E^v (\mathbf{k},m)-\omega - i\Gamma}           \Big].
\end{eqnarray}

$\textbf{P}$ is the momentum operator, $f(E^{c,v} (\mathbf{k},m)$ the Fermi-Dirac distribution function; $\Gamma$ the broadening parameter. The absorption spectrum is associated with the velocity matrix elements (the first term) and the joint density of states (the second term). The former can determine whether the inter-LL transitions are available. The velocity matrix elements, as successfully done for carbon-related materials,\,\cite{FM;PRB1994} are evaluated under the gradient approximation in the form of
\begin{eqnarray}
\Big\langle \Psi^{c} (\mathbf{k},m')
 \Big| \frac{   \hat{\mathbf{E}}\cdot \mathbf{P}   } {m_e}
 \Big| \Psi^{v}(\mathbf{k},m)    \Big\rangle
 &\cong \frac{ \mathrm{\partial}}{\mathrm{\partial}k_y  }
\Big\langle \Psi^{c} (\mathbf{k},m')  \Big| H \Big|  \Psi^{v}(\mathbf{k},m)\Big\rangle
\nonumber
\end{eqnarray}
\begin{eqnarray}
= \sum_{l,l' = 1}^{3} \sum_{m,m' = 1}^{2R_B}
\bigg (
c_{A_{m,\mathbf{k}}^{l}}^*  c_{A_{m',\mathbf{k}}^{l'}}
\frac{ \mathrm{\partial}}{\mathrm{\partial}k_y  }
 \Big\langle A_{m,\mathbf{k}}^{l} \Big| H \Big|  A_{m,\mathbf{k'}}^{l'}\Big\rangle
 \nonumber
\end{eqnarray}
\begin{eqnarray}
+ c_{A_{m,\mathbf{k}}^{l}}^*  c_{B_{m',\mathbf{k}}^{l'}}
\frac{ \mathrm{\partial}}{\mathrm{\partial}k_y  }
 \Big\langle A_{m,\mathbf{k}}^{l} \Big| H \Big|  B_{m,\mathbf{k'}}^{l'}\Big\rangle
 \nonumber
\end{eqnarray}
\begin{eqnarray}
+ c_{B_{m,\mathbf{k}}^{l}}^*  c_{A_{m',\mathbf{k}}^{l'}}
\frac{ \mathrm{\partial}}{\mathrm{\partial}k_y  }
 \Big\langle B_{m,\mathbf{k}}^{l} \Big| H \Big|  A_{m,\mathbf{k'}}^{l'}\Big\rangle
  \nonumber
\end{eqnarray}
\begin{eqnarray}
+ c_{B_{m,\mathbf{k}}^{l}}^*  c_{B_{m',\mathbf{k}}^{l'}}
\frac{ \mathrm{\partial}}{\mathrm{\partial}k_y  }
 \Big\langle B_{m,\mathbf{k}}^{l} \Big| H \Big|  B_{m,\mathbf{k'}}^{l'}\Big\rangle
\bigg ).
\end{eqnarray}

In this approximation, we do not need to really do the inner product of the left side in Eq. (4); that is, the tight-binding functions of the 2$p_z$ orbital are not included in the calculation, but only amplitudes are sufficient in the right-hand side of Eq. (4). It means that, the subenvelope functions can be used to calculated the magneto-absorption spectra. The similar theoretical framework is available in understanding the quantum Hall conductivities. In general, we can fully understand the critical factors purely due to the characteristics of LLs, e.g., monolayer graphene presents many symmetric delta-function-like absorption peaks with uniform intensity in the magneto-absorption spectrum.\,\cite{CY;IOP2017}

\medskip
\par

In general, there are two kinds of theoretical models to study the magnetic quantization phenomena, namely, the low-energy elective-mass approximation and tight-binding model. Concerning the low-energy perturbation method,\,\cite{MK;PRL2001, MT;PRB2005} the zero-field Hamiltonian matrix elements are expanded about the high-symmetry points (e.g., $\bf{K}$ point in graphene). And then, the magnetic quantization is further made from an approximate Hamiltonian matrix. That is to say, the zero-field and magnetic Hamiltonian matrices have the same dimension. However, some interlayer hopping integrals in layered graphene will create much difficulty in the study of magnetic quantization. Some of them are usually ignored in the effective-mass approximation. Consequently, certain unique and diverse magnetic quantization phenomena are lost by using this method, e.g., the anticrossing phenomenon in ABC-stacked trilayer graphene, and the extra magneto-absorption selection rules.\,\cite{XW;PCCP2017} In general, this perturbation method cannot deal with the low-symmetry systems with multi-constant energy loops. For AB-bt bilayer silicene, maybe it is impossible to solve for the low-lying energy bands with the use of a low-energy expansion method; that is, the effective-mass model is not suitable for expanding the low-energy electronic states from the $\bf{K}$ and $\bf{T}$ points simultaneously. This model becomes too cumbersome  to generate the further magnetic quantization. So, it is very difficult to comprehend the LLs, being attributed to the unique Hamiltonian in bilayer silicene. It is in sharp contrast with the monolayer silicene case.
\medskip
\par

Within the tight-binding method, the magnetic phases due to the vector potential are included in the calculations. In the previous studies,\,\cite{DR;PRB1976,  MK;AP1985} this model is developed using the $\vec{\bf{k}}$-scheme, but not the $\vec{\bf{r}}$-scheme. This is, the magnetic states are built from the original electronic states in the first Brillouin zone (the hexagonal Brillouin zone in graphene). However, it is not suitable to present the main features of LL wavefunctions (oscillatory distribution in real space with localization centers). Explicitly, the subenvelope functions could not be identified as the LL wavefunctions since they are only the random distributions. This scheme is very difficult to deal with the essential properties under spatially modulated/non-uniform magnetic field, the modulated electric field, and the composite magnetic and electric fields, e.g., the magneto-optical properties and magneto-Coulomb excitations.
For the generalized tight-binding model used in this study, the calculations are based on the sublattices in an enlarged unit cell in the real space. Under a perpendicular magnetic field, the Hamiltonian matrix becomes very huge so that we need to arrange it in a band-like form. The magnetic Hamiltonian is dependent on $k_x$  and $k_y$ in the reduced first Brillouin zone. Moreover, LL energies are fully degenerate in this Brillouin zone. The LL degeneracy is $D=\frac{\vec{\bf B} .  S_{hex}} {h/2e} \approx \frac{32000}{B_z}$. In addition, the original hexagonal first Brillouin zone is changed into a small rectangular one. The degenerate ($k_x$,$k_y$) states in the reduced first Brillouin zone only make the same contribution for any physical properties. Therefore, we choose the ($k_x$ = 0,$k_y$ = 0) state to study the magneto-electronic properties of bilayer silicene.

\section{Results and Discussion }

\subsection{Electronic structure}

\begin{figure}
\centering
{\includegraphics[width=0.7\linewidth]{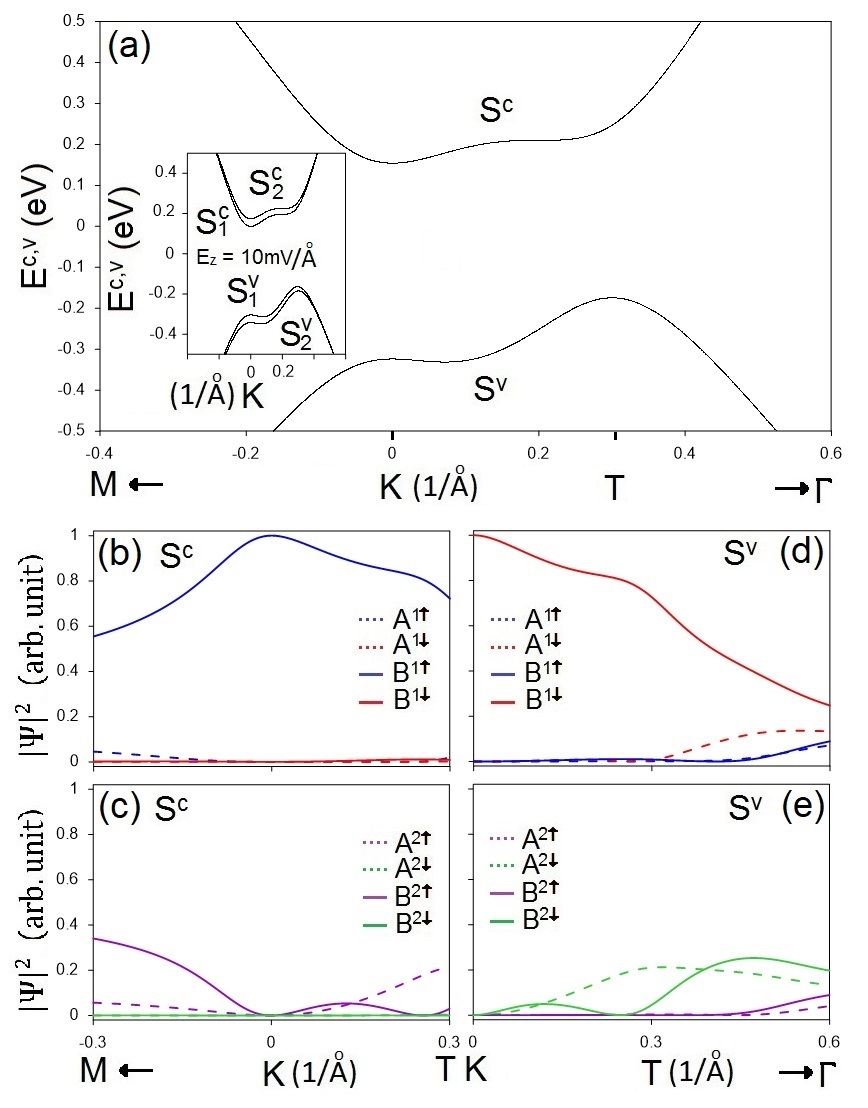}}
\caption{(Color online)  The first pair of conduction and valence energy bands (a) at $E_z=0$, the state probabilities for
the conduction band [(b), (c)] and the valence band [(d), (e)]. Also shown in the inset of (a) are
$E_z$-split energy bands.}
\label{Fig2}
\end{figure}

AB-bt bilayer silicene exhibits feature-rich band structure due to its buckled lattice,
complex intra- and inter-layer atomic interactions, and significant SOCs. There exist two
pairs of conduction and valence bands. This work will mainly address the electronic properties
of the low-lying energy bands, as clearly indicated in Fig.\,\ref{Fig2}($a$). The conduction and
valence bands display an asymmetric energy spectrum about the Fermi level (${E_F=0}$), strong
energy dispersions, a highly anisotropic behavior, and a spin-dependent double degeneracy (the
spin-up- and spin-down-dominated degenerate states discussed with respect to
Figs.\,\ref{Fig2}($b$)-\ref{Fig2}($e$)). The conduction-band valley is initiated from the $\bf{K}$ point and
presents a special shoulder-like structure along the $\bf{K}$-$\bf{\Gamma}$ direction  in the range of 0.2 eV${<E^c<0.22}$ eV.
On the other hand, the valence-state valley is built from the $\bf{T}$ point between the $\bf{K}$ and $\bf{\Gamma}$ points; furthermore, the unusual energy spectrum, with an extreme $\bf{K}$ point, is revealed along the $\bf{TK}$ direction at ${E^v\sim\,-0.33}$ eV.
There exists a noticeable indirect gap of $0.3\,$eV, corresponding to the highest occupied
state at the $\bf{T}$ point and the lowest unoccupied state at the $\bf{K}$ point. This is in sharp contrast
with the zero-gap band structures of bilayer graphene.\,\cite{YKH;SR2014} These special properties
leave footprints in the different magneto-electronic properties discussed in the next Subsection.
\medskip

The state probabilities on the distinct sublattices, $A^{1,2}$ and $B^{1,2}$ with spin-up and spin-down
configurations ($\uparrow$ and $\downarrow$), could provide mutual dependence among them, as clearly
presented in Figs.\,\ref{Fig2}($b$)-\ref{Fig2}($e$). The doubly-degenerate states have identical wave
functions under  interchange of $(B^1_{\uparrow}, B^1_{\downarrow}, A^1_{\uparrow}, A^1_{\downarrow})$
and $(B^2_{\downarrow}, B^2_{\uparrow}, A^2_{\downarrow}, A^2_{\uparrow})$, and each one exhibits very
strong sublattice, spin and wave vector dependence. The conduction states are dominated by the ${B^1_{\uparrow}}$
sublattice, especially for the full dominance at the $\bf{K}$ point (the solid blue curve in \,\ref{Fig2}($b$)). The
${B^2_{\uparrow}}$ sublattice also makes some important contributions to the $\bf{K}$-valley states (the solid purple
curve in Fig.\ref{Fig2}($c$)). As for the $\bf{T}$-valley valence states, the ${B^1_{\downarrow}}$ sublattice shows
strong dominance (the red curve in Fig.\ref{Fig2}($d$)), accompanied by partial contribution from the
${B^2_{\downarrow}}$ sublattice (the green curve in Fig.\ref{Fig2}($e$)). The ${A^{1,2}}$ sublattices do
not show the dominating features, and the dominant ${B^{1,2}}$ sublattices are expected to determine largely
the quantum modes of the magnetic LLs.

\medskip
\par

A uniform perpendicular electric field can drastically modify the electronic properties of bilayer
silicene, mainly owing to the destruction of the ${z=0}$ mirror symmetry.
The field also breaks down the spin-dependent state degeneracy and changes considerably the energy gap, leading to a dramatic
transformation of the band structure.
It is noticed that the effects due to an external electric field are revealed through the different Coulomb potentials
on each sublattice which depends on the atom heights. Such a field does not change either the geometric structure or the hopping integrals.
Each conduction/valence band is split into a pair of energy
subbands, as denoted by $S^{c}_{1,2}$ and $S^{v}_{1,2}$ in the inset of \,\ref{Fig2}($a$) for $E_z = 10\,$ mV/\AA.
The first and second conduction (valence) subbands are characterized, respectively, by the dominant
${B^2_{\downarrow}}$ and  ${B^1_{\uparrow}}$ (${B^1_{\downarrow}}$ and  ${B^2_{\uparrow}}$)
sublattices, for which the former is relatively close to the Fermi level. Such a feature will be magnified
in the $E_z$-enriched LL energy spectra, as discussed below. Moreover, the sizable band gap is easily
tuned by the external electric field. With further increase of the field strength, the $S^{c}_{1}$ and
$S^{v}_{1}$ energy subbands will be pushed down to the Fermi level, while the opposite is true for the
$S^{c}_{2}$ and $S^{v}_{2}$ ones. Apparently, the band gap is reduced and then vanishes beyond a
critical field ($E_z$ = 106 meV), and a semiconductor-semimetal transition occurs at higher electric fields.
This makes bilayer silicene extremely useful for electronic device applications.

\subsection{The quantized Landau levels}

\begin{figure}
\center
{\includegraphics[width=0.8\linewidth]{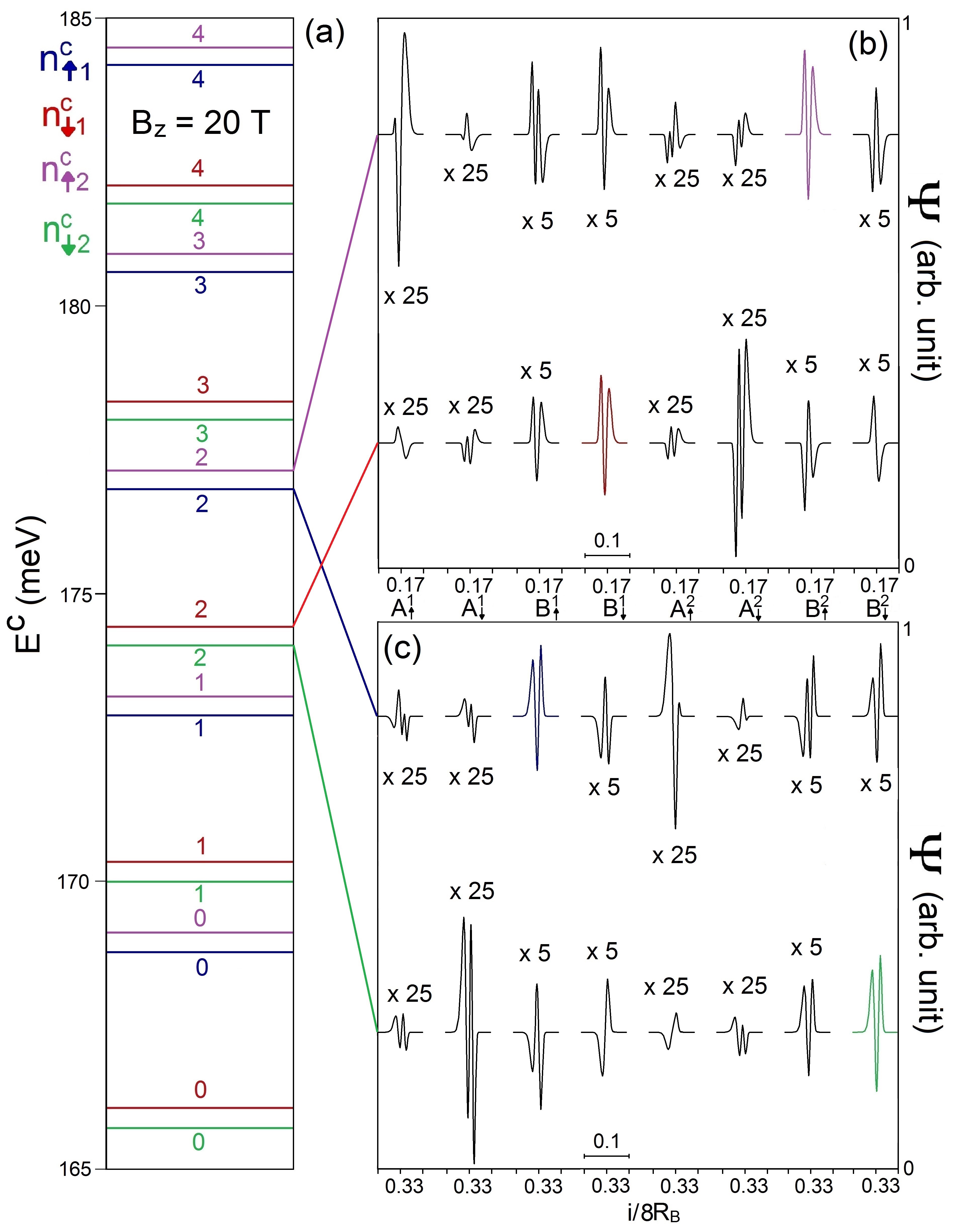}}
\caption{(Color online)  At $B_z = 20$ T, the low-lying conduction LL energies (a); the spatial amplitude distributions
for the $B^2_{\uparrow}$- and $B^1_{\downarrow}$-dominated LLs (b) and the $B^1_{\uparrow}$- and
$B^2_{\downarrow}$-dominated LLs (c). The numbers in the abscissa represent the quantum numbers of those Landau levels. They are determined by the number of zero points of the spatial distributions on the dominating sublattices.}
\label{Fig3}
\end{figure}

\begin{figure}
\centering
{\includegraphics[width=0.8\linewidth]{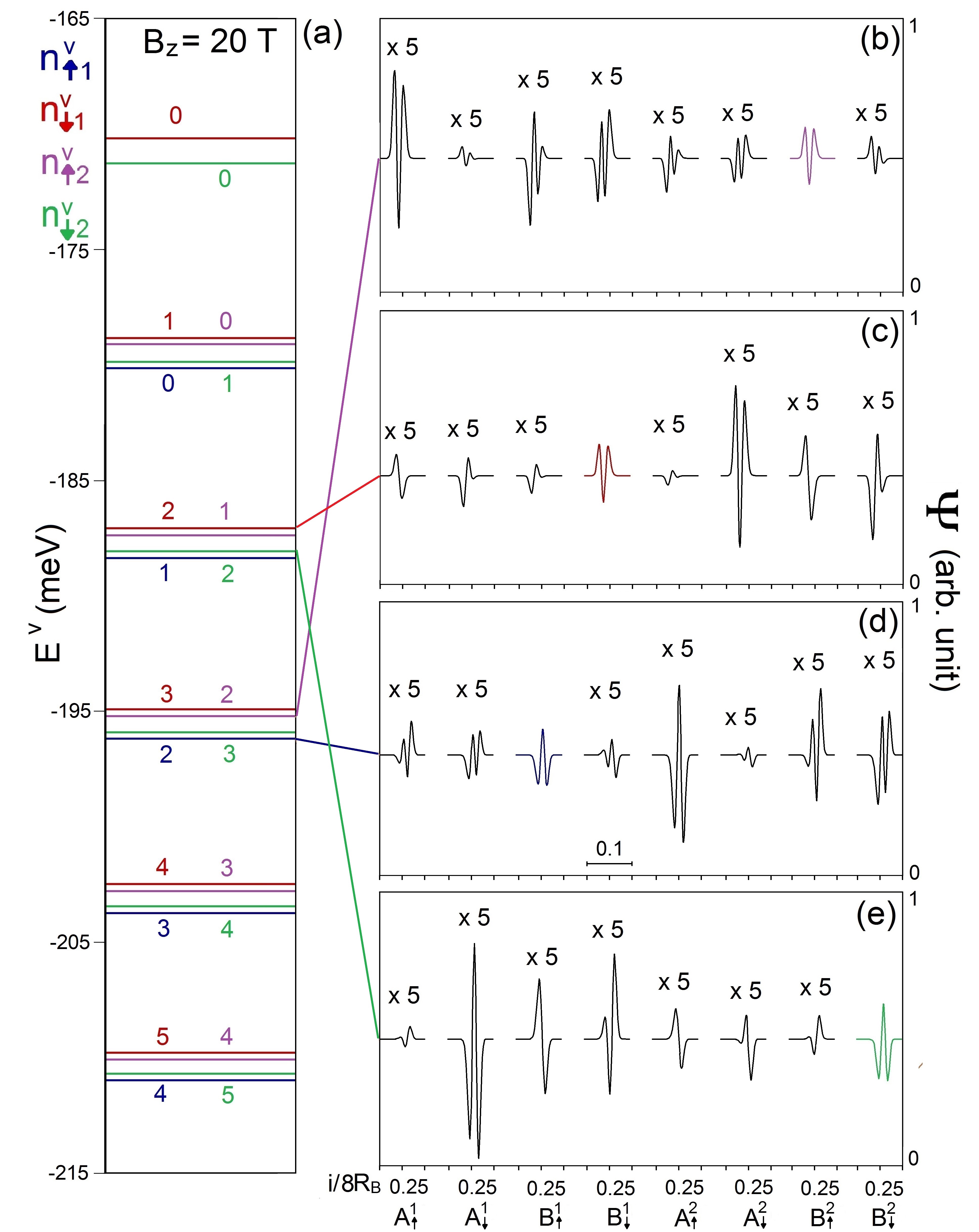}}
\caption{(Color online)  At $B_z = 20$ T, the low-energy valence LLs (a); the spatial wave functions for
the $B^2_{\uparrow}$- (b), $B^1_{\downarrow}$- (c), $B^1_{\uparrow}$- (d) and $B^2_{\downarrow}$-dominated
(e) LLs.}
\label{Fig4}
\end{figure}

The low-lying LLs possess rich characteristics due to the buckled structure, strong inter-layer atomic
interactions, and sizable SOCs. All the LLs are degenerate in the reduced first Brillouin zone with
${|k_x|\le\,2\pi / aR_B}$ and ${|k_y|\le\,2\pi /\sqrt{3}a}$ (an area of $4\pi^2 / \sqrt{3}a^2 R_B$). The
${(k_x=0,k_y=0)}$ magnetic state is sufficient for understanding the main behaviors of magnetic quantization.
The conduction and valence LLs, corresponding to the magnetic quantization of their electronic states,
respectively, near the $\bf{K}$ (or $\bf{K^{\prime}}$) and $\bf{T}$ points, are asymmetric with respect to $E_F$ = 0
(Figs.\,\ref{Fig3} and \ref{Fig4}). They are reduced to doubly degenerate under the broken equivalence
of the (B$^1$, B$^2$) sublattices and  spin-up and spin-down configurations. The non-degenerate conduction
(valence) states are localized at $1/6$ and $2/6$ ($1/4$) of the expanded unit cell. One should note
that the conduction [valence] LLs are degenerate at the (1/6 and 4/6) and (2/6 and 5/6) [(1/4 and 3/4)] localization
centers.\,\cite{YKH;SR2014}

\medskip
\par

As for the low-lying conduction LLs, there exist four non-degenerate states for a chosen
quantum number, as illustrated by blue, red, green and purple colors in Fig.\,\ref{Fig3}($a$) for
$B_z=20$ T. They are distinguished by LL wave functions based on the ${3p_z-}$ and spin-dependent
sub-envelope functions on the separate sublattices. The spatial distributions are similar to the oscillation
modes of a harmonic oscillator. For example, the ${n=2}$ LLs acquire well-behaved spatial oscillations with
two nodes in the dominant sublattices, in which the ${B^2_{\uparrow}}$ and ${B^1_{\downarrow}}$-dominated states
(the ${B^1_{\uparrow}}$ and the ${B^2_{\downarrow}}$-dominated ones) correspond to the 1/6 (2/6) localization
center by the purple and red curves in Fig.\,\ref{Fig3}($b$)  (the blue and green curves in Fig.\,\ref{Fig3}($c$)).
In parallel with $B^{1}_{\uparrow}$, $B^{1}_{\downarrow}$, $B^{2}_{\uparrow}$ and $B^{2}_{\downarrow}$, the
sublattice- and spin-dominated LLs could be classified into four subgroups:
$n^{c}_{\uparrow 1}$, $n^{c}_{\downarrow 1}$, $n^{c}_{\uparrow 2}$ and $n^{c}_{\downarrow 2}$.
Here, $c$ and $v$ represent the conduction and valence LLs, respectively; 1/2 is the dominating $B_1$/$B_2$ sublattice and $\uparrow$ $(\downarrow)$ indicates the dominating spin-up (spin-down) state of the subgroup. With this notation, we find it convenient to show the critical mechanisms related to LLs, such as, the inter-LL optical transition (Fig. 9), and the inter-LL Coulomb scattering.\,\cite{JYW;ACSN2011}
Each LL subgroup exhibits normal ordering for increasing state energy, i.e., $n^c$ increases with $E^c$.
However, the LL energy spacing decreases with increasing $n^c$.   There exists a very narrow energy spacing
between the LLs at $1/6$ (purple/red curve) and $2/6$ (blue/green curve) centers under similar spin
dominance, as seen from Fig.\,\ref{Fig3}($a$). For example, the sublattice-dependent energy splitting
between the $n^{c}_{\uparrow 2}$ = 0 (1/6 center) and $n^{c}_{\uparrow 1}$ = 0 (2/6 center) is only about
$0.5\,$meV. This is much smaller than the spin-induced energy splitting, e.g., ${\sim\,3}\,$meV for the
$n^{c}_{\uparrow 1}$ = 0 and $n^{c}_{\downarrow 1}$ = 0 LLs (or ${n^c_{\uparrow 2}}$ and
${n^c_{\downarrow 2}}$ ones). The above-mentioned features also appear in the valence LLs, as clearly
indicated in Figs.\,\ref{Fig4}($a$)-\ref{Fig4}($e$). There are four valence LL subgroups,
$n^v_{\uparrow 1}$, $n^v_{\downarrow 1}$ $n^v_{\uparrow 2}$ $n^v_{\downarrow 2}$
(Fig.\,\ref{Fig4}($a$)). The ${B^1_{\downarrow}}$-,  ${B^2_{\downarrow}}$-,  ${B^2_{\uparrow}}$-, and
${B^1_{\uparrow}}$-dominated LLs (Figs.\,\ref{Fig4}($b$)-\,\ref{Fig4}($e$)) shows the same 1/4 localization center.
The non-equivalence of the $B^1$ and $B^2$ sublattices and the absence of spin-state degeneracy are responsible
for these unique LLs.
\medskip

\begin{figure}
\centering
{\includegraphics[width=0.7\linewidth]{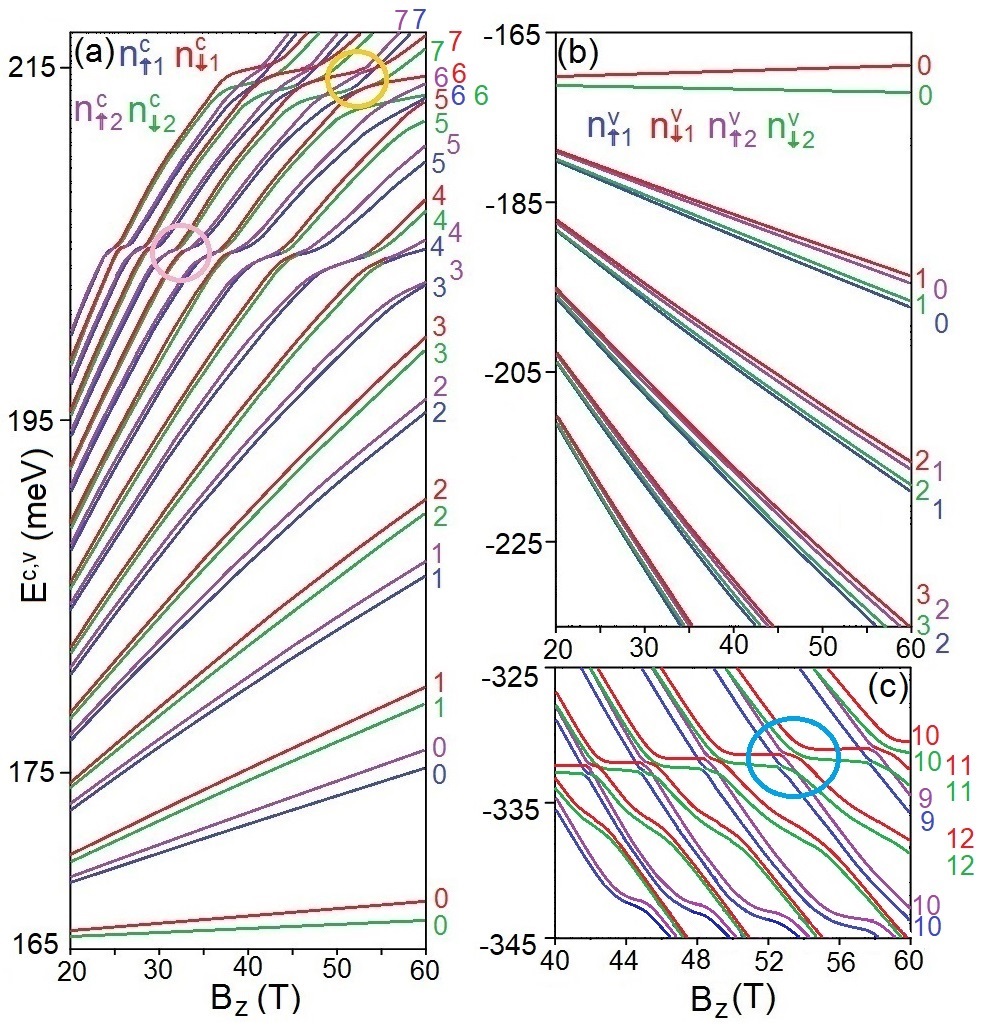}}
\caption{(Color online)   The $B_z$-dependent LL energy spectrum: the blue, red, purple, and green curves
are the different subgroups of conduction (a) and valence (b)-(c) LLs based on the dominant
${B^1_{\uparrow}}$, ${B^1_{\downarrow}}$, ${B^2_{\uparrow}}$ and ${B^2_{\downarrow}}$ sublattices, respectively.}
\label{Fig5}
\end{figure}

\medskip
\par

The $B_z$-dependent LL energy spectrum, which is identified from the unique sublattice- and spin-dominated
LL wave functions in bilayer AB-bt silicene, is critical in comprehending the diverse quantization phenomena.
As for the conduction-LL energy spectrum, four distinct subgroups exhibit similar magnetic field dependence,
as clearly shown in Fig. \,\ref{Fig5}($a$). In the range of $165\,$meV${\le\,E^c\le\,195}\,$meV,  the
small-$n^c$ LL energies have monotonic/almost linear $B_z$-dependence and the normal ordering among four
subgroups. However, the abnormal behaviors, i.e.,  unusual field dependence and LL anti-crossings, show up
frequently at higher energies. The ${n^c_{\uparrow 1}}$ and ${n^c_{\downarrow 2}}$ LLs (blue and green curves)
anti-cross with each other  within a specific $B_z$ range, and so do the ${n^c_{\downarrow 1}}$ and
${n^c_{\uparrow 2}}$ LLs (red and purple curves), as marked by the pink and yellow circles for the ${n^c=7}$ LLs.
Therefore, two kinds of inter-subgroup anti-crossings appear frequently for the specific quantum modes.
These anti-crossings clearly indicate that the wave functions of the perturbed LLs include the  main and side modes,
but not a single mode. Furthermore, such modes change substantially as the field strength is varied
(discussed later with respect to Fig.\,\ref{Fig7}).\,\cite{CYL;PCCP2015, YKH;SR2014} Similarly, there exists
a simple relationship between the valence LL energies and the field strength for
$-300\,$meV${\le\,E^v\le\,-165}\,$meV. The deeper-energy spectrum presents two kinds of inter-subgroup LL
anti-crossings arising from neighboring LLs, i.e., [${n^v_{\uparrow 1}}$ $and$  ${n^v_{\downarrow 2}+2}$]
(blue and green curves in Fig.\,\ref{Fig5}($c$)) and [${n^v_{\uparrow 2}}$ $and$  ${n^v_{\downarrow 1}+2}$]
(purple and red curves). Here, the $n^v =10$ ($n^v =12$) LLs acquire a side mode of 12 (10), leading to the
anti-crossings for (${n^v_{\uparrow 1}}=10$ and  ${n^v_{\downarrow 2}}=12$) and (${n^v_{\uparrow 2}}=10$ and
${n^v_{\downarrow 1}}=12$) LLs, as indicated by the blue circle in Fig.\,\ref{Fig5}($c$).
According to the Wigner-von Neuman non-crossing rule, two multi-mode (single-mode) LLs avoid crossing each other
when they simultaneously possess certain identical modes (nonidentical mode) with comparable amplitudes on the
specific sublattices. The details of LL anti-crossings  have been shown in Figs.\,\ref{Fig3} and \ref{Fig4}. The
rich and unique LL energy spectra are closely related to the magnetic quantization of the unusual conduction
and valence bands.

\medskip

\begin{figure}
\centering
{\includegraphics[width=0.8\linewidth]{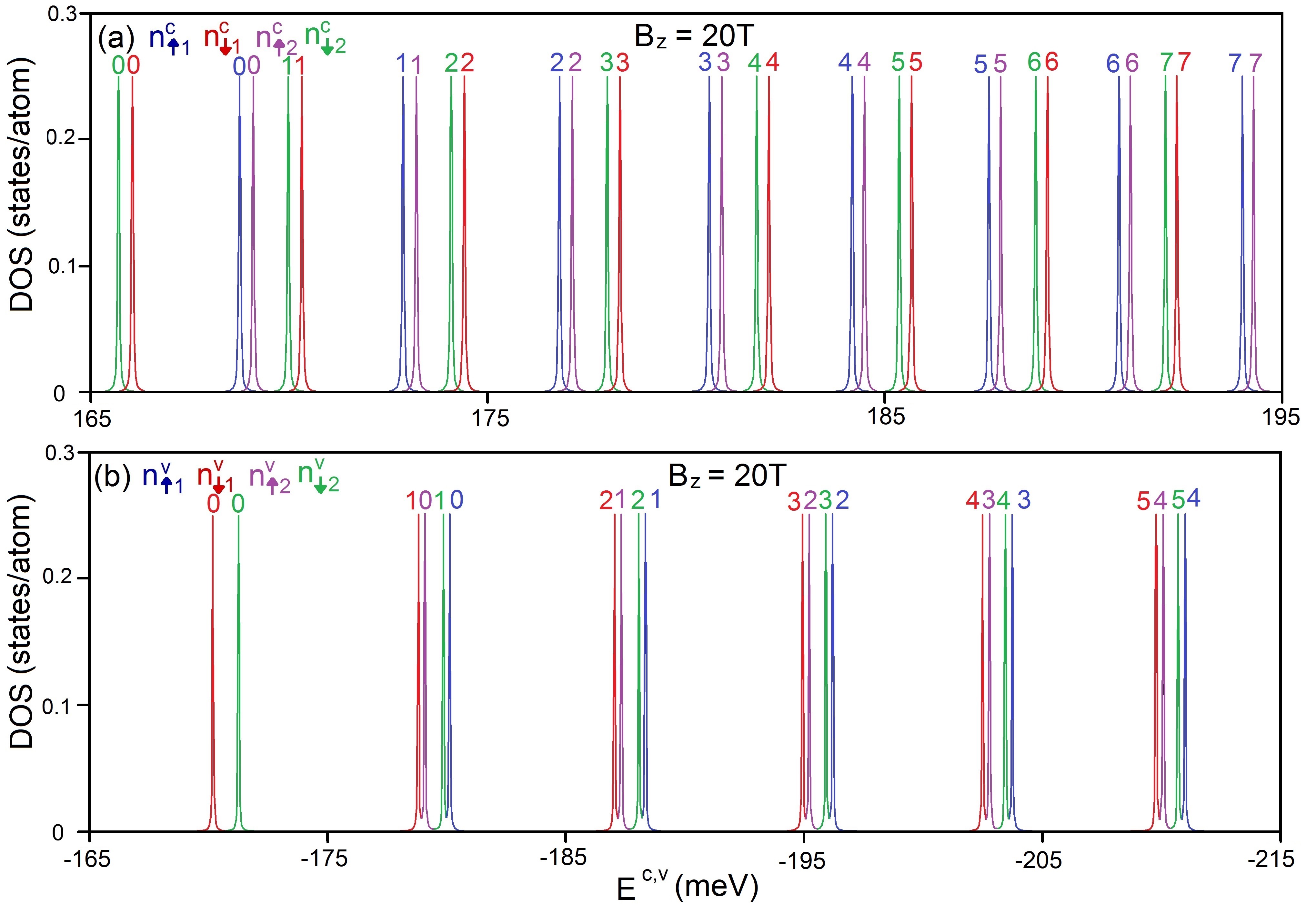}}
\caption{(Color online)  The density-of-states of the conduction (a) and valence (b) LLs under a
magnetic field of $B_z$ = 20 T.}
\label{Fig6}
\end{figure}

\medskip
\par

The rich sublattice- and spin-dominated LL energy spectra can be seen from the DOS, as shown
in Figs.\,\ref{Fig6}($a$) and \ref{Fig6}($b$). Four subgroups of conduction and valence LLs are
clearly displayed in the form of $\delta$-function-like peaks. The low-energy DOS exhibits a uniform
peak height in the presence of the same LL degeneracy and the absence of LL crossings. In each subgroup,
the spacing between two neighboring peaks gradually decreases with increasing energy. Moreover, the
subgroup-dependent DOS peaks appear under a specific ordering. The DOS of the conduction LLs  presents
the ordering of $n^c_{\downarrow2}$-, $n^c_{\downarrow 1}$-, $n^c_{\uparrow 1}$- and $n^c_{\uparrow 2}$-dominated
peaks (green, red, blue and purple in Fig.\,\ref{Fig6}($a$)), while the sequence for the DOS of the
valence LLs  is $n^v_{\downarrow 1}$-, $n^v_{\uparrow 2}$-, $n^v_{\downarrow 2}$- and $n^v_{\uparrow 1}$-induced
peaks (red, purple, green and blue peaks in Fig.\,\ref{Fig6}($b$)), except for the ${n^v_{\downarrow 1}=0}$ and
${n^v_{\downarrow 2}=0}$ initial peaks. The well-behaved LL DOS peaks disappear at higher energy as a
result of the frequent anti-crossings/crossings (Figs.\,\ref{Fig5}($a$) and \,\ref{Fig5}($c$)).

\medskip
\par

The main features of the DOS peaks, covering the structure, height, number and energy, could be verified by scanning tunneling
spectroscopy (STS).\,\cite{GHL;PRL2009, YJS;N2010, WXW;PRB2015,GMR;NP2011, LJY;PRB2016,LJY;PRB2015,GHL;NP2007}
To our knowledge, STS measurements have been successfully used to identify the magneto-electronic energy for layered graphene.
The LL energies in monolayer\,\cite{GHL;PRL2009, YJS;N2010, WXW;PRB2015} and AB bilayer
graphenes\,\cite{GMR;NP2011, LJY;PRB2016} are confirmed with the square-root and linear dependence on
$B_z$, respectively. The coexistence of $\sqrt{B_z}$ and linear $B_z$-dependent LL energies in ABA-stacked
trilayer graphene,\,\cite{LJY;PRB2015} as well as the 3D and 2D characteristics of the Landau subbands in
AB-stacked graphite\,\cite{GHL;NP2007}, have also been examined. Moreover, the LL crossings and anti-crossings
 phenomena can also be verified by measuring the shift of the plateaus in the quantum Hall effect (QHE)
due to the increase in the LL degeneracy,\,\cite{TT;NP2008} while the latter can be examined by identifying
the non-monotonic changes in the structures, energies, intensities and numbers of the DOS peaks.\,\cite{MKF;PRL2010}
Finally, the unusual magneto-electronic properties in AB-bt bilayer silicene could be further verified by STS,
including the normal and unusual $B_z$-dependence, the sublattice- and spin-dominated four subgroups of LLs,
as well as the special LL splittings, crossings and anti-crossings.

\subsection{The electric field-enriched Landau levels}

\begin{figure}
\centering
{\includegraphics[width=0.85\linewidth]{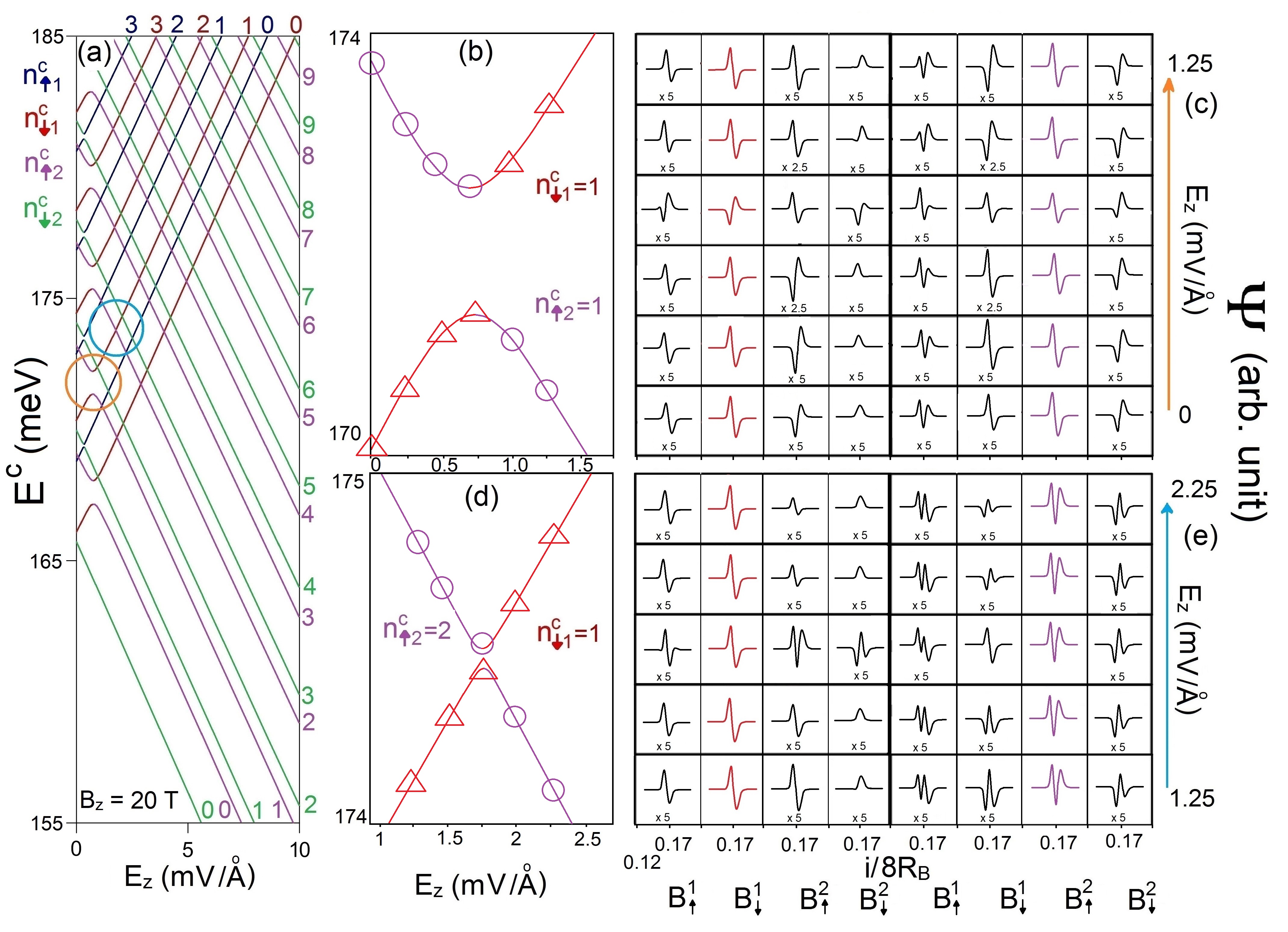}}
\caption{(Color online)  The $E_z$-dependent conduction-LL energy spectrum (a) at ${B_z=20}$ T; the [${n^c_{\downarrow\,1}=1}$ $\&$ ${n^c_{\uparrow\,2}=1}$] and [${n^c_{\downarrow\,1}=1}$ $\&$ ${n^c_{\uparrow\,2}=2}$] anti-crossings (b), corresponding to
the evolutions of wavefunctions in (c) and (e), respectively.}
\label{Fig7}
\end{figure}

\begin{figure}
\centering
{\includegraphics[width=0.85\linewidth]{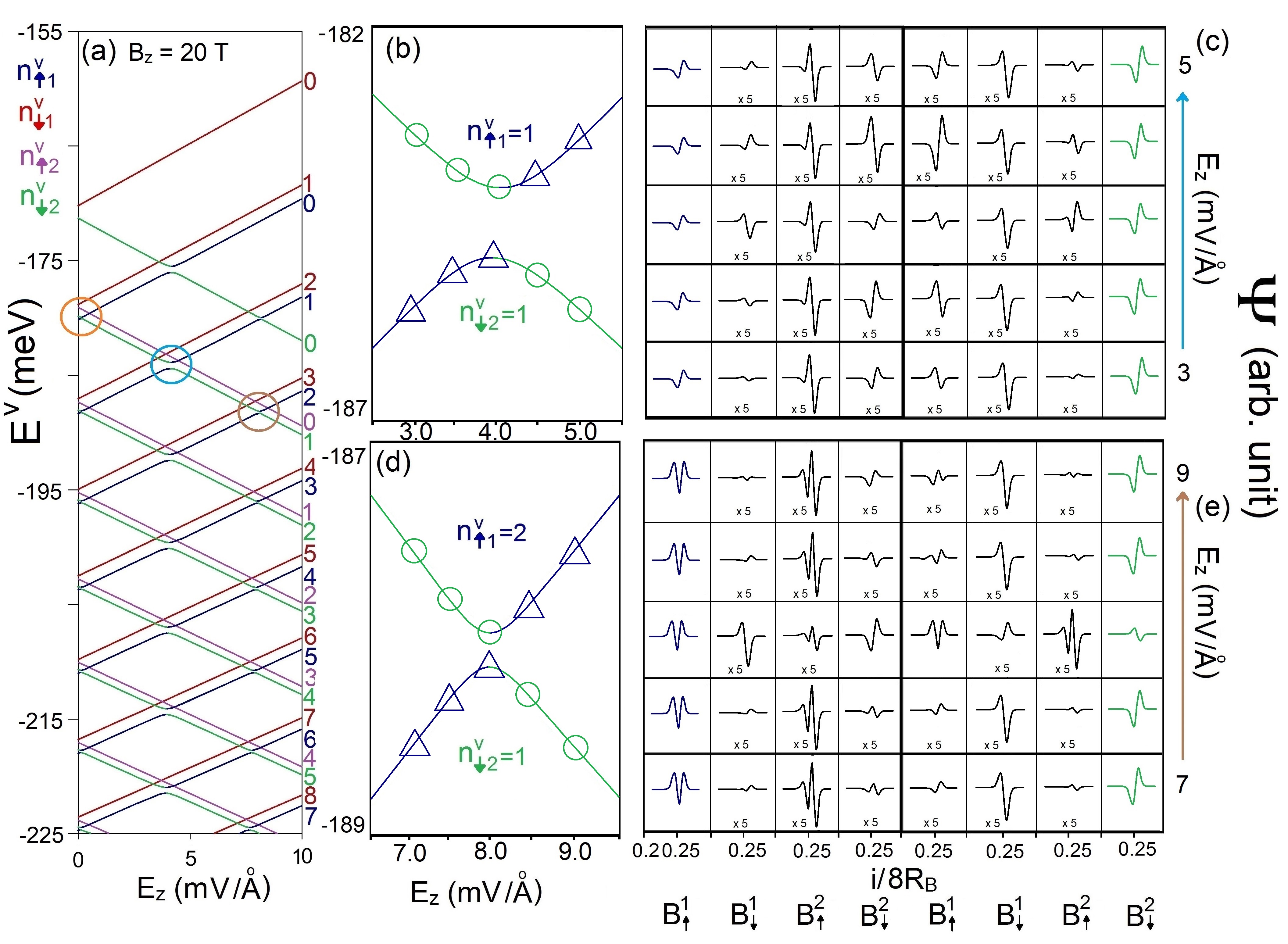}}
\caption{(Color online)  The $E_z$-dependent valence-LL energy spectrum (a) at ${B_z=20}$ T; the [${n^v_{\downarrow\,2}=1}$ $\&$ ${n^v_{\uparrow\,1}=1}$] and [${n^v_{\downarrow\,2}=1}$ $\&$ ${n^v_{\uparrow\,1}=2}$] anti-crossings (b), corresponding to
the evolutions of wavefunctions in (c) and (e), respectively.}
\label{Fig8}
\end{figure}

The magneto-electronic properties can be remarkably diversified by applying an external electric field.
In the presence of composite electric and magnetic fields, the LL energies, which belong to
[$n^{c,v}_{\uparrow\,1}$ and $n^{c,v}_{\downarrow\,1}$] along with  [$n^{c,v}_{\uparrow\,2}$ and
 $n^{c,v}_{\downarrow\,2}$] subgroups, respectively, increase or decrease with the electric
field strength, as clearly shown in Figs.\,\ref{Fig7}($a$) and \,\ref{Fig8}($a$) by the [blue and red curves]
and the [purple and green curves]. The $E_z$ dependence of these subgroups is opposite for the ${B^1}$ and
$B^2$-dominated LL energies. This clearly indicates $E_z$-enhanced sublattice non-equivalence by means of
distinct Coulomb potential energies. There exist many LL anti-crossings and crossings in the $E_z$-dependent
energy spectra, in which the former are induced by the neighboring [$n^{c,v}_{\uparrow\,1}$ and
$n^{c,v}_{\downarrow\,2}$] LLs (blue and green curves) and  [$n^{c,v}_{\downarrow\,1}$ and the
$n^{c,v}_{\uparrow\,2}$] ones (red \$ purple curves). However, such behaviors are absent between the
$n^{c,v}_{\uparrow\,1}$ and $n^{c,v}_{\downarrow\,1}$ subgroups as well as between the
$n^{c,v}_{\uparrow\,2}$ and $n^{c,v}_{\downarrow\,2}$ subgroups. That is, the spin-related LL energy spacing
is weakly affected by $E_z$.

\medskip
\par

\medskip
\par

Here, the LL anti-crossings are worthy of close examination. In the $E_z$-dependent energy spectrum, the
inter-subgroup anti-crossings are only related to two neighboring LLs with quantum number difference
$\Delta n=0,\,\pm\,1,\,\pm\,2$. These electronic transitions are either difficult to observe or changed
to crossing behaviors. The conduction LLs anti-cross each other for [${n^c_{\downarrow, 1}}$ and
${n^c_{\uparrow, 2}}$], [${n^c_{\downarrow, 1}}$ and  ${n^c_{\uparrow, 2}\,+1}$] as well as
 [${n^c_{\uparrow, 1}}$ and ${n^c_{\downarrow, 2}+1}$] as shown in Fig.\,\ref{Fig7}($a$).
The valence LLs behave similarly, as seen in Fig.\,\ref{Fig8}($a$). The ${n^c_{\downarrow\,1}\,=1}$ LL
is chosen as an example for understanding the evolution of the  wave functions as the field strength
is increased (red curve in Fig.\,\ref{Fig7}($b$)). In this case, the ${n^c_{\downarrow\,1}\,=1}$ LL has its
first anti-crossing with the ${n^c_{\uparrow\,2}\,=1}$ LL as $E_z$ increases initially from zero (red and
purple curves). They remain in their states for (${n=1}$) on the ${B^1_{\downarrow}}$ and ${B^2_{\uparrow}}$
sublattices, as displayed in Fig.\,\ref{Fig7}($c$). However, one of two amplitudes (red) gives rise to phase
switching during the anti-crossing, e.g., those at $E_z=0$ and $0.75\,$mV/\AA. Moreover, the wave functions of
the $n^c_{\downarrow 1}=1$ and $n^c_{\uparrow 2}=1$ LLs on the $B^2_{\uparrow}$ and $B^1_{\downarrow}$ sublattices,
respectively, are greatly enhanced near the anti-crossing center and acquire comparable amplitudes
commensurate with the dominating ones (right- and left-hand sides black curves and by the red and purple
curves). With further increase of $E_z$, the ${n^c_{\downarrow\,1}\,=1}$ and ${n^c_{\uparrow\,2}\,=2}$ LLs
encounter another weak anti-crossing at ${E_z\sim\,1.75}\,$mV/\AA in Fig.\,\ref{Fig7}($d$). In this case, the
latter one on the $B^2_{\uparrow}$ sublattice is mapped directly onto the former wave function on the
$B^2_{\uparrow}$ sublattice and vice versa in Fig.\,\ref{Fig7}($e$)). The above-mentioned LL wave functions
agree with the  Wigner-von Neuman non-crossing rule. Similar anti-crossings can be found for valence LLs in
Fig.\,\ref{Fig8}($a$). For example, the ${n^v_{\downarrow\,2}\,=1}$ LL anti-crosses with the ${n^v_{\uparrow\,1}\,=1}$
and ${n^v_{\uparrow\,1}\,=2}$ LLs at lower and higher electric fields in Figs.\,\ref{Fig8}($b$) and
\,\ref{Fig8}($d$)), respectively, as can be identified from the drastic changes of amplitude in
Figs.\,\ref{Fig8}($c$) and \,\ref{Fig8}($e$) or even the transformation of oscillating modes on the
${B^1_{\uparrow}}$ and ${B^2_{\downarrow}}$ sublattices.

\medskip
\par

The AB-bt bilayer silicene sharply contrasts with AB-stacked bilayer graphene with
respect to  band structure and LLs. The former and latter, respectively, belong to an indirect-gap
semiconductor and a semimetal. The graphene structure   has two pairs of valence and conduction bands,
with monotonic energy dispersions, a weak anisotropy and a small overlap near the $\bf{K}$ point.
Electronic states are independent of the spin configurations in the absence of SOCs. They remain doubly
degenerate for the spin degree of freedom under any external fields. The neighboring states are
magnetically quantized into well-behaved LLs with specific nodes in the oscillating probability
distribution. For each $({k_x,k_y})$ state, the LLs are eight-fold degenerate as a result of the
equivalent sublattices and spin degeneracy. The conduction/valence LLs cannot be classified into four
LL subgroups, in which they only present the regular $B_z$-dependent energy spectra without the
anti-crossing behaviors. Moreover, a perpendicular electric field can lead to a semimetal-semiconductor
phase transition, but not the semiconductor-semimetal transition associated with the split energy bands.
For silicene, the electric field also leads to a lifting of the degenerate $\bf{K}$ and $\bf{K^{\prime}}$ valleys
for the magnetic LL states. There exist four degenerate LLs with frequent anti-crossings in the
$E_z$-dependent energy spectra.  The important differences between bilayer silicene and graphene highlight the
main features of the spin- and sublattice-dependent energy bands and LLs, demonstrating directly the
distinct geometric structures, atomic interactions and SOCs.

\subsection{The optical conductivities}

The feature-rich band structure and LLs are directly reflected in the unusual optical properties in bilayer silicene. The zero-field and magneto optical conductivities are shown in Figs. 9(a)-(b). In the absence of external fields, the low-lying absorption spectrum mainly exhibits two kinds of special structures: discontinuities and logarithmic divergences (Fig. 9(a)). There exist two shoulder-like structures at 0.42 and 0.48 eV's, and a sharp symmetric peak at 0.52 eV which are, respectively, corresponding to the vertical transitions of extreme and saddle points in the energy-wave-vector space. These optical structures can be verified by the infrared reflection spectroscopies and absorption spectroscopies.\,\cite{ZQ;NP2008, LM;PRB2008}

\medskip
\par

The magneto-absorption spectrum exhibits a lot of single, double and twin non-uniform delta-function-like peaks (Fig. 9(b)). Each symmetric peak is marked by $\omega_{n^v,n^c}$, in which $n^v$ and $n^c$ indicate the quantum numbers of the unoccupied valence and the occupied conduction LLs, respectively. It is very difficult to identify the absorption peaks because they belong to the vertical transitions of the multi-mode LLs. Such transitions do not follow a specific selection rule. When a conduction/valence LL possesses a sufficiently large quantum number, its spatial distribution will be extended along the x-axis so that it overlap with that of the valence/conduction LL. This leads to the inter-LL optical transitions between those LLs. The intensities of absorption peaks strongly depend on the overlapping relation of the initial and final LLs. The threshold peak (optical gap), which is identified as $n_{1\downarrow}^v = 0$ $\rightarrow$ $n_{1\downarrow}^c = 12$, is located at about 0.39 eV for $B_z$ = 40 T (arrow in Fig. 9(b)). This optical gap is much larger than the energy gap (0.3 eV) since the low-lying conduction and valence LLs are localized at different centers. As a result, the optical transitions among the low-lying conduction and valences LLs are forbidden. Moreover, the threshold peak frequency is relatively lower than that of the zero-field absorption spectrum. This is the strong evidence of the anti-crossing phenomena of the LLs quantized from the shoulder-like band structure. The former will be recovered to the latter when the phonon-assisted optical process are allowed at finite temperatures. In general, the optical gap is changed with the variation of the magnetic field strength.

\medskip
\par

\begin{figure}
\centering
{\includegraphics[width=0.85\linewidth]{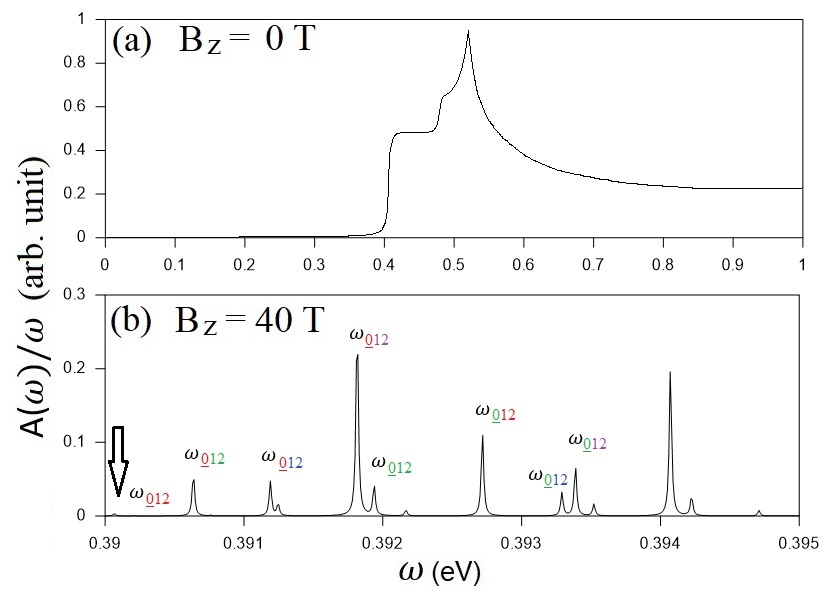}}
\caption{(Color online)  The optical conductivities of AB-bt bilayer silicene under (a) zero magnetic field and (b) $B_z$ = 40 T.}
\label{Fig9}
\end{figure}

\section{Concluding Remarks}

The magneto-electronic and optical properties of bilayer AB-bt silicene are studied by the generalized tight-binding model, combined with the dynamic Kubo formula and gradient approximation. This system presents asymmetric conduction and valence bands about ${E_F=0}$, parabolic and irregular energy dispersions, the strong anisotropy, and an observable indirect gap, leading to the rich and unique magnetic quantization. The low-lying conduction and valence LLs are doubly degenerate for each $(k_x,k_y)$ state, in which they could be classified into the sublattice- and spin-dominated four subgroups. The non-equivalence of $B^1$ and $B^2$ sublattices and spin splitting are clearly revealed in the field-dependent energy spectra. The LLs exhibit the almost linear $B_z$-dependence near $E_F$, and there exists a specific ordering among four subgroups. In general, the magnetic quantization phenomena in AB-bt bilayer silicene with complex SOCs and interlayer atomic interactions are very different from those of other 2D systems. The sublattice- and spin-dominated four groups of LLs in this system are never revealed in other 2D emergent materials up to now.
The STS measurements on LLs are very useful in examining the interplays of the complicated interlayer atomic interactions and the important SOCs. However, the abnormal energy spectra, accompanied with the frequent anti-crossings and crossings, come to exist at higher/deeper energy. The unique features in bilayer silicene are expected to be verifiable by direct experimental measurements.

\medskip
\par

The electronic properties in bilayer silicene are very sensitive to an applied electric field.
We can easily adjust the band gap, which is a crucial quantity for inducing a semiconductor-semimetal phase transition.  Furthermore, the electric field also dramatically alters the main features of the LLs,
including the spatial distribution and the anti-crossing phenomena. The $B^1$ and $B^2$-dominated LL energies present the opposite
$E_z$-dependences, so that a lot of anti-crossings and crossings appear at low field strength. The former, being associated with ${\Delta\,n=0}$, ${\pm\,1}$ and ${\pm\,2}$, are identified from the drastic changes of wavefunction amplitudes or even the mode transformation on the specific two sublattices. They mainly arise from the cooperation of the intrinsic interactions and the external fields. In short, the AB-stacked bilayer silicene and graphene quite differs from each other in band structure, LL degeneracy, field-dependent energy spectra, and anti-crossing/crossing behaviors.
We expect the utilization of bilayer silicene will bring us new opportunity in gate
controlling of magneto-quantum channel conductance which can be applied to novel designs of $Si$-based
nano-electronics and nano-devices with enhanced mobilities.

\medskip
\par
AB-bt bilayer silicene exhibits the unique and abnormal optical conductivities. In the absence of external field, the absorption spectrum shows the shoulder structures and logarithmic divergence which, respectively, come from the vertical transitions of the extreme and saddle points of energy band. The magneto-optical peaks are enriched by the inter-LL excitations among four subgroups of LLs. Such transitions do not obey a specific optical selection rules, being in sharp contrast with those in bilayer graphene. They are very sensitive to the localization centers, and their intensities depend on the LL overlap. The absorption spectrum exhibits the single-, double- or twin-peak structures with non-uniform intensities. The main features of LLs are responsible for the unusual magneto-optical conductivities.

\medskip
\par

For the complicated band structure of bilayer silicene, the magnetic quantization can only be solved by the generalized tight-binding model. This is because the geometric symmetry, the intrinsic interactions (hopping integrals and SOCs), and the external fields need to be solved by the generalized tight-binding model simultaneously. On the other hand, the effective-mass model is very difficult/impossible to solve the magnetic quantization by the low energy expansion because of distinct valleys, strong anisotropy and weakly energy dispersions in conduction and valence bands. Even when the low-energy conduction $\textbf{K}$ valley and the valence $\textbf{T}$ one can be done, it is not exact to make the magnetic quantization separately, which is inconsistent with the requirement of quantum statistics (the distribution of Fermions).

\medskip
\par

The tight-binding model is capable of directly combining with the single- and many-particle theories to explore other fundamental physical properties, such as, the magneto-optical properties, quantum Hall effect, and magneto-Coulomb excitations. We have built up a theoretical framework to investigate the critical physical properties of condensed-matter systems. The optical conductivities are studied using the dynamic Kubo formula and gradient approximation. The quantum Hall effect is explored by the employment of the static Kubo formula within the linear response theory. Moreover, the random-phase approximation needs to be modified to agree with the layer-dependent tight-binding model, and then it is available for exploring the magneto-electronic excitations in few-layer silicene systems. That is, intra- and inter-layer atomic interactions and intra- and inter-layer Coulomb interactions are taken into account simultaneously. The whole theoretical framework is very concise and meaningful.

\medskip
\par

\begin{acknowledgements}
We would like to acknowledge the financial support from the Ministry of Science and Technology of Taiwan
(R.O.C.) under Grant No. MOST 105-2112-M-017-002-MY2 and the US Air Force Office of Scientific Research (AFOSR).
\end{acknowledgements}

\newpage
 
$\textbf{References}$

\end{document}